\documentclass[preprint]{elsarticle}

\usepackage{lineno,hyperref}
\usepackage[utf8]{inputenc}
\usepackage{multirow}
\usepackage{nth}
\usepackage{url}
\usepackage{csquotes}
\usepackage{booktabs}
\usepackage[inline]{enumitem}  
\usepackage{hyperref}
\usepackage{fourier}
\usepackage{mathtools}  
\usepackage{subfigure}
\usepackage{xcolor}

\let\linenumbers\nolinenumbers\nolinenumbers

\journal{Journal of Applied Soft Computing}

\begin{document}

\begin{frontmatter}

\title{Short Text Classification Approach to Identify Child Sexual Exploitation Material}

\cortext[cor1]{E-mail addresses: \{mnab, efidf, ealeg, ralar\}@unileon.es}
\author[ULE,INCIBE]{Mhd~Wesam~Al-Nabki}
\author[ULE,INCIBE]{Eduardo~Fidalgo}
\author[ULE,INCIBE]{Enrique~Alegre}
\author[ULE,INCIBE]{Rocío~Alaiz-Rodríguez}

\address[ULE]{Department of Electrical, Systems and Automation, Universidad de León, Spain}
\address[INCIBE]{Researcher at INCIBE (Spanish National Cybersecurity Institute), León, Spain}
\address{}





\begin{abstract}
Producing or sharing Child Sexual Exploitation Material (CSEM) is a serious crime fought vigorously by Law Enforcement Agencies (LEAs). When an LEA seizes a computer from a potential producer or consumer of CSEM, they need to analyze the suspect's hard disk's files looking for pieces of evidence. However, a manual inspection of the file content looking for CSEM is a time-consuming task. In most cases, it is unfeasible in the amount of time available for the Spanish police using a search warrant. Instead of analyzing its content, another approach that can be used to speed up the process is to identify CSEM by analyzing the file names and their absolute paths. The main challenge for this task lies behind dealing with short text distorted deliberately by the owners of this material using obfuscated words and user-defined naming patterns.
This paper presents and compares two approaches based on short text classification to identify CSEM files. The first one employs two independent supervised classifiers, one for the file name and the other for the path, and their outputs are later on fused into a single score. Conversely, the second approach uses only the file name classifier to iterate over the file's absolute path. Both approaches operate at the character n-grams level, while binary and orthographic features enrich the file name representation, and a binary Logistic Regression model is used for classification. The presented file classifier achieved an average class recall of $0.98$.
This solution could be integrated into forensic tools and services to support Law Enforcement Agencies to identify CSEM without tackling every file's visual content, which is computationally much more highly demanding.

\end{abstract}


\end{frontmatter}

\linenumbers

\section{Introduction}
\label{sec:introduction}
In $2017$, the Council of the European Union (EU) prioritized cybercrimes related to Child Sexual Abuse (CSA), considering them as the most serious crimes between the years $2018$ and $2021$ \cite{child2019POLICY}. 
According to The European Police Office, Child Sexual Exploitation Material (CSEM) is defined as sexual abuse of a person under 18 years old and producing images or videos of the abuse and distributing such content online \cite{child2019Europl}. 

Darknets, such as The Onion Router (Tor)\footnote{\url{https://www.torproject.org/}} \cite{alnabki2019torank,he2019classification,alnabki2017classifying} and FreeNet\footnote{\url{https://freenetproject.org/}} \cite{levine2017statistical} and also Peer to Peer (P2P) networks, like eDonkey, \cite{panchenko2012detection,peersman2016icop} are environments where the interchange of CSEM seem to proliferate, thanks to the high level of privacy and anonymity provided to their users. These characteristics allow pedophiles to easily share CSEM far away from Law Enforcement Agencies (LEAs) monitoring. It is worth mentioning that during the COVID19 outbreak, Interpol has reported a significant increase in exchanging CSA material in P2P and Darknet networks as well as online gaming and messaging applications \cite{covidInterpol}.

CSEM producers and consumers might save this content on their local computer machines, at least temporarily. When an LEA inspects a home to analyze a  suspect's computers, a police agent reviews the files in the investigated hard drive, trying to determine whether or not the suspected of pedophilia has stored CSEM in the computer \cite{Gangwar_2017_CSAVideo}. This process needs to be accomplished in a limited time and as accurately as possible \cite{Chaves_2019_SpeedAccuracyFaceDetector}. This work aims to build a File Classifier (FC) that decides whether a given file is related to CSEM or not according to its name and absolute path. The FC will not tackle the content as other modules, which are out of this paper's scope. Hence, FC will act as a preliminary filter in a CSEM detection pipeline.

Building an automatic FC is a challenging task due to several reasons.
Firstly, a binary supervised algorithm requires training samples of Non-CSEM and CSEM files. However, there are no publicly available datasets of the latter class, and crawling samples from a P2P network or the Darknet is illegal \cite{garcia2018textile}. Therefore, only CSEM file names obtained legally, i.e., provided by LEAs, could be used.
Secondly, a file name typically is a text of small length, which leads to a sparse representation of the samples because we have a massive number of features, while an instance is only represented with a few of them.
Finally, CSEM producers or consumers tend to invent a personalized file name style to create their vocabulary, abbreviations, and acronyms to circumvent detection tools, using a personalized obfuscated writing style. For example, in a sample named ``\textit{!!!!yoB0yXX}", the exclamation marks refer to the age of a boy, and the letter \textit{O} is replaced by the number zero. Hence, most likely, this file is related to the abuse of a four years old boy. 
It worth mentioning that the last two challenges, i.e. the lack of context and the deliberate distortion of the text, make it more difficult to build or to use pre-trained language models, such as word2vec \cite{mikolov2013distributed} and GloVe \cite{pennington2014glove}, and contextualized word representations, like Bidirectional Encoder Representations from Transformers (BERT) \cite{luo2018active} and Embeddings from Language Models (ELMO) \cite{peters-etal-2018-deep}.
Nevertheless, high accuracy may not be achievable as some of these resources carry minimal information. This would play a key role in filtering those files with a higher probability of being CSEM and facilitating the analysts’ work that otherwise would be unfeasible.

A small body of research was focused on the problem of identifying CSEM via their file names. The most recent works are the research of Pereira et al. \cite{pereira2020metadata} and Al-Nabki et al. \cite{alnabki2020filenameclassification} where they experimented with different deep learning and machine learning algorithms to build a supervised classifier for the files. 

Unlike the common strategies that use file names only, this paper attempts to dive further to incorporate their absolute path in parallel. The approach of using both pieces of information, i.e. the name and the path, has been presented for the first time by Pereira et al. \cite{pereira2020metadata} when they train a single classifier to classify the absolute file path, including the file name. In contrast, the File Classifier (FC) we propose in this paper uses dedicated classifiers for each component, a File Path Classifier (FPC) to classify the absolute paths and a File Name Classifier (FNC) for file names. Their outputs are fused into a single score. This design will prevent the classification decision from skew to the absolute file path's content only, which typically occupies most of the text. 

We counted on our previous work for the FNC design \cite {alnabki2020filenameclassification} but after extending the file name representation and using a bigger dataset. Furthermore, this paper elaborates on the FPC and demonstrates how its output was integrated with the FNC.

\begin{figure}[htp]
\centering
\includegraphics[scale=0.7]{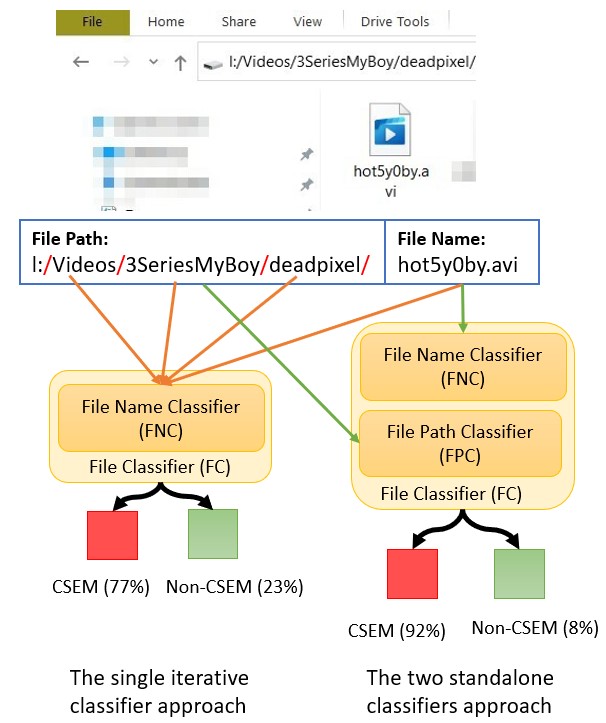}
\caption{Two classification approaches to classify files. The structure shown to the right of the figure demonstrates how the two standalone classifiers approach works. In this configuration, FNC and FPC classify the file name and file path, respectively. Then, their outputs are fused to determine the file category. While the structure shown on the left illustrates how the single iterative classifier approach works. In this configuration, FNC is used to iterate over the path's directories and the name of the file.}
\label{fig:fnc_clf}
\end{figure}

The file name and the file path could complement each other when any of them carries a CSEM pattern. Nevertheless, this approach will not be advantageous when neither sources exist, such as a file named only with numbers and located in the root folder. 
We propose two approaches to build the FC (see Fig. \ref{fig:fnc_clf}). 
The first one uses two standalone classification models, one for the FNC and another for the FPC. The outputs of these two classifiers are fused into a single output.
The other approach employs the FNC only to classify the file name and the path. It iterates over the absolute path along with the file name, and whenever the FNC detects a CSEM name within the path, it reports the file as CSEM.

The main contributions of this paper are summarized as follows.

\begin{itemize}

\item We propose a framework for classifying files into CSEM or safe material based on the fusion of the output of two supervised classifiers, which uses file names and their absolute paths. 

\item We extend the text of the file names by appending two additional intermediate representations suited for the task of CSEM detection. The first one is a novel binary representation, which distinguishes character blocks from non-character ones. The second is an orthographic feature that captures the variation in the types of file name characters. To the best of our knowledge, the orthographic feature has not been used before to code file names for text classification, but for named entity recognition tasks \cite{al2020improving, limsopatham2016learning}. 

\item We build a dataset with $5.9$M and $890$K unique file paths and file names samples, respectively. To be the best of our knowledge, this is the largest dataset used for classifying CSEM using file names and paths.

\item We apply our framework into a real-case application: CSEM detection. We also introduce our framework into a practical forensic tool that could support the task of CSEM detection to the LEA worldwide.

\end{itemize}

The rest of the paper is organized as follows: Section \ref{sec: related-work} presents the related work. Section \ref{sec: methodology} describes the proposed classification methodology. Next, Section \ref{sec: dataset} explains how the datasets of the FNC and the FPC are created and what are their main features. Then, in Section \ref{sec: eval}, we describe the experimentation performed and discuss the results obtained. Finally, Section \ref{sec: conclusion} presents the conclusions by pointing to our future research.

\section{Related Work}
\label{sec: related-work}
The use of filename classification in recognizing CSEM has not received much attention despite its efficacy in identifying potential forensic evidence. To the best of our knowledge, only a few research papers have been published in recent years.

To begin with, Panchenko et al. \citep{panchenko2012detection} attempted to normalize file names using Short Message Service (SMS) normalization techniques proposed by Beaufort et al. \cite{beaufort2010hybrid}. With the normalized text, they trained a Support Vector Machine (SVM) classifier and obtained an accuracy of $96.97\%$ on their dataset. Peersman et al. \cite{peersman2014icop} proposed a framework called iCOP to detect CSEM on P2P networks. The first stage of their classification pipeline was a dictionary-based filter that was constructed manually and held CSEM keywords. They used character n-gram of size two to four to capture more features about the file name and a binary SVM as classifier. Afterward, in their recent work \cite{peersman2016icop}, Peersman et al. used a similar representation but benchmarked more classifiers, like SVM and Naive Bayes (NB). Due to the lack of a public dataset for this task, they evaluated their proposal on a custom dataset as well, and they observed that the SVM classifier could identify CSEM file names with a recall rate of $0.43$. Al-Nabki et al. \cite{alnabki2020filenameclassification} compared the use of machine learning classifiers, such as SVM and LR, that use character n-gram with Term Frequency-Inverse Document Frequency (TF-IDF), versus deep learning classifiers that depend on Convolutional Neural Network (CNN). Specifically, they adopted two CNN models developed by Zhang et al. \cite{zhang2015character} and Kim et al. \cite{kim2016character}. The model of Zhang et al. was the best benchmarked CNN-based classifier and obtained an F1 score of $0.85$, while the machine learning approach using LR classifier scored a slightly lower F1 score of $0.84$. The major difference was their processing time on a CPU machine where the latter was by far quicker than the former one. Pereira et al. \cite{pereira2020metadata} compared several machine learning and deep learning models to classify files using the file names and paths. They conducted the experiments on a dataset of $1,010,000$ file paths from $55,312$ unique storage systems provided by Project VIC International. Similar to the conclusion of Al-Nabki et al. \cite{alnabki2020filenameclassification}, they found out that the CNN character-based model proposed by Zhang et al. \cite{zhang2015character} achieves the best recall rate of $0.94$

The problem of CSEM identification through file names could be approximated to a wider research topic, such as short text classification \cite{sriram2010short,sun2012short,shang2013feature,rana2014news,alsmadi2018term,vskrlj2020tax2vec}, and in particular, news headlines classification and Twitter posts classification.

The news headlines classification task attempts to group news articles based on their titles, in which the title typically is made up of a few words. Rana et al. \cite{rana2014news} proposed a pipeline of three stages: data pre-processing, text representation, and classification. In the data pre-processing step, the text was tokenized into words, and spaces replaced special characters, stop words were removed, and the text was stemmed. For the text representation, the authors used TF-IDF, Information Gain (IG) \cite{shang2013feature}, and Boolean Weight (BW) \cite{chouchoulas1999rough}. Finally, in the classification stage, Rana et al. explored NB \cite{kim2006some}, SVM \cite{Joachims1998Text}, K-Nearest Neighbor (KNN) \cite{hotho2005brief}, and Decision Trees (DT) \cite{safavian1991survey}. However, the core difference between our problem and news headlines classification is that the latter has high-quality input text, where the punctuation marks are maintained correctly, and there are no misspelled words.

Classifying tweets of Twitter would also fall under the umbrella of short text classification as the common length of a tweet is $33$ characters, while the maximum number of characters is $280$ \cite{perez_2019}.
Furthermore, the quality of the text could be low in comparison to the news headlines problem, and it might contain abbreviations to save space or misspell some words \cite{perez_2019}. Imran et al. \cite{imran2016cross} pre-processed the tweets by removing hyperlinks, mentions, and stop words. Then, they used the N-grams and IG techniques for feature extraction and a Random Forest (RF) classifier \cite{Breiman2001}. Chen et al. \cite{Chen2018} proposed a framework to identify cyberbullying on Twitter. For text representation, they compared pre-trained language models, like Word2Vec \cite {mikolov2013distributed} and GloVe \cite{pennington2014glove}, with traditional text encoding techniques, such as TF-IDF, and they realized a decline in the performance when embedding-based were used. For classification, they compared traditional machine learning classifiers like LR and SVM with deep learning classifiers, like Long Short-Term Memory (LSTM) \cite{lee-dernoncourt-2016-sequential} and Convolutional Neural Network (CNN) \cite{zhang2015character}. Carnevale et al. \cite{carnevale2020investigating} proposed an algorithm to classify noisy and low-quality text generated from critical patients’ posts on Twitter. The authors employ n-gram with TF-IDF for feature extraction and benchmark two classifiers, SVM and NB, to compare their performance on the task.

Furthermore, the problem of file path classification could be treated as a branch of URL classification since both share similar characteristics in terms of the structure and the use of concatenated words. This topic has been investigated widely by many researchers \cite{singh2017online,zouina2017novel,rajalakshmi2018naive,trevisan2019robust,sahingoz2019machine, sharma2020malicious}. Sahingoz et al. \cite{sahingoz2019machine} used a URL classification approach to identify phishing websites through their addresses. Sahingoz et al. explore various features extracted manually from the URL, and they used them to benchmark several machine learning classifiers, such as DT, SVM, and RF. Trevisan et al. \cite {trevisan2019robust} examined the use of Generative Adversarial Neural Networks (GANs) to classify four classes of URLs given its ability to cope with the lack of training samples problem. 

\section{Methodology}
\label{sec: methodology}
This section introduces two approaches for designing the File Classifier (FC) in order to identify CSEM. The first one involves two standalone classifiers, one for the file name and another for the file path, and the outputs of these two classifiers are fused to a single value that represents the prediction confidence. The second approach is to build a single classifier for the file names. Since the path consists of a sequence of file names, a file name classifier can iterate over the sub-directory names starting from the root directory to the file name. Finally, the prediction confidences of the sub-directories are fused. Both approaches have a typical machine learning design that consists of three main stages \cite{alnabki2017classifying}: text pre-processor, feature extractor, and a classifier. In the following, we elaborate on each approach in detail.

\subsection{Two Standalone Classifiers Approach}
In the following, we present two classifiers, a File Name Classifier (FNC) and a File Path Classifier (FPC). Each classifier has its dataset for training and testing.

\subsubsection{File Name Classifier}
\label{sec:file_name_clf}
The FNC presented in this paper attempts to enhance the previous implementations explored in the literature by: 1) enhancing the file name representation and 2) training on a larger and more representative dataset (Section \ref{lab:fnc_dataset}).

\paragraph{File Name Pre-processing}
To extend the representation of a file name, we enrich the original file name by three text pre-processing techniques, whereas their outputs are concatenated to form an input to the next component in the classification pipeline. Table \ref{tab: samplerepresentation} shows how two samples are prepared.

The text pre-processing function replaces special characters and numbers by \textit{\#} and \textit{\$}, respectively, to reduce the sparsity of the features. For instance, a file named ``\textit{!!!!yoB0yXX}" will be transformed to ``\textit{\#\#\#\#yoB\$yXX}". This technique was beneficial to detect and eliminate duplicated samples that differ only by parts of their names. For example, a folder in a seized computer could have more than $100$ images named IMG01.png, IMG02.png, ... IMG100.png, whereas all these names are repeating the same information of IMG\#\#.png. 
However, we realized that there was a significant portion of misclassified files because they did not contain explicit text indicating its category. As an example, a sequence of random characters and digits, like ``\textit{DJ4MD9F34SE45EX8CH85YO}" and ``\textit{QBDD35HMF93DF5TVH4TD}". Exploring the two and three grams of the former example, we observe grams like "8YO", "SE", and "EX", which might seem like indicators of suspiciousness. However, we can realize that this name is not suspicious when looking at the complete file name. 
This problem motivated us to enrich the representation of the file names by adding a new binary representation. This technique replaces each block of characters by one and by zero for anything else, i.e. digits and special characters. Hence, the above mentioned examples will be represented as ``\textit{1010101010101}" and ``\textit{101010101.jpg}", while an example such as ``\textit{Hot3YoGirlOnBeach}" will be mapped to ``\textit{101}". This representation can distinguish oscillations between characters and non-characters block in the input text. 

Additionally, we added an orthographic feature as it showed a significant improvement in our previous work for the noisy Named Entity Recognition task \cite{al2020improving}. This feature maps capital letters, small letters, numbers, and special characters to unique tokens as ``C", ``c", ``N", and ``P", respectively. As an example, a file name like ``\textit{!!!!yoB0yXX}" will be mapped to ``\textit{PPPPccCNcCC}".

\begin{table}[htp]
\centering
\caption{Example of preprocessing and tokenizing a file name with two, three, four, and five grams}
\label{tab: samplerepresentation}
\begin{tabular}{ll}
\hline
\multicolumn{1}{c}{\textbf{Representation levels}} & \multicolumn{1}{c}{\textbf{Encoded text}} \\ \hline
Input text & Hot3YoGirlOnBeach \\
Digit \& special char. & Hot\$YoGirlOnBeach \\
Binary representation & 101 \\
Orthographic representation & CccNCcCcccCcCcccc \\
Output text & \begin{tabular}[c]{@{}l@{}}Hot\$YoGirlOnBeach 101\\ CccNCcCcccCcCcccc\end{tabular} \\ \hline

Input text & !!!!yoB0yXX \\
Digit \& special char. & \#\#\#\#yoB\$yXX \\
Binary representation & 0101 \\
Orthographic representation & PPPPccCNcCC \\
Output text & \begin{tabular}[c]{@{}l@{}}\#\#\#\#yoB\$yXX 0101\\PPPPccCNcCC\end{tabular} \\ \hline

Input text & FG5F44GDSdfs234DG \\
Digit \& special char. & FG\$\$F\$\$GDSdfs\$\$DG \\
Binary representation & 1010101 \\
Orthographic representation & CCNCNNCCCcccNNNCC \\
Output text & \begin{tabular}[c]{@{}l@{}}FG\$\$F\$\$GDSdfs\$\$DG 1010101 \\ CCNCNNCCCcccNNNCC\end{tabular} \\ \hline
\end{tabular}
\end{table}

\paragraph{File Name Feature Extraction}
\label{lab: fnc_feat_ext}
Finding an adequate representation of the text is a crucial step in a classification pipeline. For this work, we used character n-gram to extract all the patterns of two to five consecutive characters of an input file name, which builds a set of tokens. Then, we apply the well-known TF-IDF technique \cite{aizawa2003information} since it gives higher weight scores to grams whose frequency is higher in a few file names and, at the same time, decreases the weight of grams that frequently occur in many files. This way, it overcomes the issue of misspelled words or personalized naming style in file names. Table \ref{tab: n-gram_rep} shows an example of two to five grams of a file name ``\textit{!!!!yoB0yXX}". Furthermore, to discard noisy tokens, we set thresholds for the minimum and the maximum term frequency. 

\begin{table}[htp]
\centering
\caption{Example of pre-processing and tokenizing a file name with two to five grams}
\label{tab: n-gram_rep}
\resizebox{\linewidth}{!}{
\begin{tabular}{lc}
\hline
\multicolumn{1}{c}{\textbf{\begin{tabular}[c]{@{}c@{}}Original\\ File Name\end{tabular}}} & \textbf{!!!!yoB0yXX} \\ \hline
\textbf{Pre-processing} & \textbf{\#\#\#\#yoB\$yXX} \\
2-grams & \#\#, \#\#, \#\#, \#y, yo, oB, B\$, \$y, yX, XX \\
3-grams & \#\#\#, \#\#\#, \#\#y, \#yo, yoB, oB\$, B\$y, \$yX, yXX \\
4-grams & \#\#\#\#, \#\#\#y, \#\#yo, \#yoB, yoB\$, oB\$y, B\$yX, \$yXX \\
5-grams & \#\#\#\#y, \#\#\#yo, \#\#yoB, \#yoB\$, yoB\$y, oB\$yX, B\$yXX \\ \hline
\textbf{Binary representation} & \textbf{0101} \\ \hline
2-grams & 01, 10, 01 \\
3-grams & 010, 101 \\
4-grams & 0101 \\
5-grams & - \\ \hline
\textbf{Orthographic representation} & \textbf{PPPPccCNcCC} \\ \hline
2-grams & PP, PP, PP, Pc, cc, cC, CN, Nc, cC, CC \\
3-grams & PPP, PPP, PPc, Pcc, ccC, cCN, CNc, NcC, cCC \\
4-grams & PPPP, PPPc, PPcc, PccC, ccCN, cCNc, CNcC, NcCC \\
5-grams & PPPPc, PPPcc, PPccC, PccCN, ccCNc, cCNcC, CNcCC \\ \hline
\end{tabular}
}
\end{table}

\paragraph{File Name Classification}
After having the features extracted, we use them for training the FNC. Based on previous research \cite{alnabki2020filenameclassification} and considering both classification performance and execution time, we use Logistic Regression.

\subsubsection{File Path Classifier}
\label{sec:file_path_clf}
The FPC is a supervised binary classifier to decide whether a given file's absolute path is CSEM related or not. The FPC consists of the following three components:

\paragraph{File Path Pre-processing}
Only the file paths are pre-processed at this stage since the FNC already handled the file names. Initially, the path is converted into a string by replacing the slash sign ($/$) with space. Next, we replace special characters and digits by $\#$ and $\$$, respectively. Finally, using the regular expression library, we split the text by capital letters if exist. 
Table \ref{tab: FPC_preprocessing} illustrates the pre-processing procedure applied to two samples. 

\begin{table}[htp]
\centering
\caption{Stages of the path pre-processing procedure}
\label{tab: FPC_preprocessing}
\resizebox{\linewidth}{!}{
\begin{tabular}{ll}
\hline
\multicolumn{1}{c}{\textbf{\begin{tabular}[c]{@{}c@{}}Pre-processing\\ Stage\end{tabular}}} & \multicolumn{1}{c}{\textbf{Text}} \\ \hline
Input Text & /Home/Downloads/MyImages/MadridTrip\_05\_05\_2020/IMG\_1.png \\
\begin{tabular}[c]{@{}l@{}}Remove \\ file name\end{tabular} & /Home/Downloads/MyImages/MadridTrip\_05\_05\_2020 \\
Replace (/) sign & Home Downloads MyImages MadridTrip\_05\_05\_2020 \\
Replacing special char. & Home Downloads MyImages MadridTrip\#\$\$\#\$\$\#\$\$\$\$ \\
Split on capital letter & Home Downloads My Images Madrid Trip\#\$\$\#\$\$\#\$\$\$\$ \\ \hline
Input Text & l:/Videos/Voyeur/3SeriesMyBoy/deadpixel/deadpixel--ro10.avi \\
\begin{tabular}[c]{@{}l@{}}Remove \\ file name\end{tabular} & l:/Videos/Voyeur/3SeriesMyBoy/deadpixel/ \\
Replace (/) sign & l: Videos Voyeur 3SeriesMyBoy deadpixel \\
Replacing special char. & l\# Videos Voyeur \#SeriesMyBoy deadpixel \\
Split on capital letter & l\# Videos Voyeur \#Series My Boy deadpixel \\ \hline
\end{tabular}
}
\end{table}

\paragraph{File Path Feature Extraction}
The problem of path classification is similar to the file name classification. In both cases, we could not use pre-train models because most of the text will be out of vocabulary and will not be represented properly. For this reason, we used the same feature extraction technique we used for the FNC, i.e. using n-gram technique, between two to five grams, that works on the character level. We applied it along with TF-IDF algorithm, as described in Section \ref{lab: fnc_feat_ext}.

\paragraph{File Path Classification}
After having the features extracted from the file paths, we use them for training a binary supervised Logistic Regression classifier, which will identify CSEM paths from the regular ones.

\subsubsection{Fusing File Name and File Path Classifiers}
This section aims to present how we aggregate the prediction of the two classifiers, the FNC and the FPC, into one prediction value. The desired fusion strategy must be sensitive to potential CSEM, either in the file name or the file path. Hence, our fusion strategy returns the result of the classifier, which has the highest CSEM confidence. For example, for a given sample $x$, the FNC predicts it is CSEM with $20\%$ confidence and $80\%$ otherwise, while the FPC predicts it is  CSEM with $40\%$ and Non-CSEM with $60\%$. In this case, the FPC confidence for the CSEM is higher than the FNC's confidence, and therefore the result of the FPC will be the final output of the fusion. Formally, Eq. \ref{eq:clf_fusion} explains the following procedure.

\begin{equation}
\label{eq:clf_fusion}
  FC(x)=\left\{
  \begin{array}{@{}ll@{}}
    FNC (x), & \text{if}\ FNC(x)_{CSEM}>FPC(x)_{CSEM} \\
    FPC (x), & \text{otherwise}
  \end{array}\right.
\end{equation} 

where $FC(x)$ refers to the classification result of a sample $x$, and the $FNC(x)_{CSEM}$ and $FPC(x)_{CSEM}$ refer to the classifier confidence regarding the CSEM class.

\subsection{Single Iterative Classifier Approach}
Typically, the absolute path of a file is made up of a sequence of folder names. This approach considers that each folder is a standalone file name, and it uses the previously implemented FNC (presented in Section \ref{sec:file_name_clf}) to classify it. Therefore, if an entry path has $N$ sub-directories, including the file name, the FNC will be called $N$ times and classify $N$ entries. If any of these $N$ entries were reported as CSEM, the complete path is considered CSEM. Otherwise, the entry is considered as non-CSEM. Unlike the majority voting approach, this technique is highly sensitive to any suspicious sub-directory name mentioned in the input path.
The prediction complexity of this approach is proportional to the depth of the absolute path. Hence, for $M$ samples and each has $N$ sub-directory, the complexity would be $O(N\times M)$.

\section{Dataset Construction}
\label{sec: dataset}
Recalling this paper's objective is to build a classifier to identify suspicious files via their metadata, i.e., name and path, without tackling their content. Previously, we presented a reduced dataset of file names \cite{alnabki2020filenameclassification}. In this work, we present a more representative and larger dataset of file names provided by the LEAs and present another one for the file paths.

Typically, each file is associated with a path to indicate its location on the hard disk. Spanish LEA shared with us a list of file names extracted from seized hard drives. These files have two patterns (i) files with representative text, i.e., explicit CSEM-related text, but without paths, as if they were extracted from the root directory of a hard disk, and (ii) files with unrepresentative names, such as IMG\_01.png, IMG\_02.png, etc., but with their absolute paths. 

Another motivation to split up on these two themes is that a unique path could contain hundreds of files, resulting in hundreds of file name samples. Table \ref{tab: path_examples} shows five file names that refer to two unique paths.

Furthermore, we noticed that the lack of explicit CSEM-related words distinguishes the paths. Instead, considering the words of the whole path sub-directory at once may lead to suspicious content of that path. To illustrate this, Table \ref{tab: path_examples} gives two unique path samples. In the first example, the word ``Sarah" standalone or ``Silver Starlets" are not CSEM-related, but their existence with other directories named ``Starlets", ``skirt", and the number five ( the last directory of the first example) could be an indicator of a sequence of photos for a 5-years old girl dressing a pink skirt.

\begin{table}[htp]
\centering
\caption{Examples of file paths dataset along with their corresponding file names}
\label{tab: path_examples}
\resizebox{\linewidth}{!}{
\begin{tabular}{ll}
\hline
\multicolumn{1}{c}{\textbf{File Path}} & \multicolumn{1}{c}{\textbf{File Name}} \\ \hline
l:\textbackslash{}Modeling\textbackslash{}Silver Models\textbackslash{}Sarah\textbackslash{}Silver Starlets\textbackslash{}Pink skirt 5\textbackslash{} & pinkskirt-1-028.JPG \\
l:\textbackslash{}Modeling\textbackslash{}Silver Models\textbackslash{}Sarah\textbackslash{}Silver Starlets\textbackslash{}Pink skirt 5\textbackslash{} & pinkskirt-1-029.JPG \\
l:\textbackslash{}Modeling\textbackslash{}Silver Models\textbackslash{}Sarah\textbackslash{}Silver Starlets\textbackslash{}Pink skirt 5\textbackslash{} & pinkskirt-1-030.JPG \\
l:\textbackslash{}Modeling\textbackslash{}BD Modeling\textbackslash{}Charming Models\textbackslash{} & GX8E5675.jpg \\
l:\textbackslash{}Modeling\textbackslash{}BD Modeling\textbackslash{}Charming Models\textbackslash{} & GX8E5676.jpg \\ \hline
\end{tabular}
}
\end{table}

\subsection{File Name Dataset}
\label{lab:fnc_dataset}
For the negative class, i.e. the safe files, we used a dataset published by the National Software Reference Library (NSRL)\footnote{\url{https://www.nist.gov/software-quality-group/about-nsrl/nsrl-introduction}} that contains more than $32$ million file names. We selected an initial subset of $800,000$ Non-CSEM examples, resulting in $537,807$ after applying the pre-processing step.
Regarding the CSEM class, we collected these examples thanks to the collaboration between the Spanish National Cybersecurity Institute (INCIBE)\footnote{In Spanish, it stands for the Instituto Nacional de Ciberseguridad de España} and the Spanish LEAs. This latter provided us with a list with dumps of hard disks seized from criminals' computers. The list had $90,000$ CSEM samples. However, after pre-processing them, the number decreased to $37,648$ unique instances.

\subsection{File Path Dataset}
Similar to the file name classifier, the file path classifier has two classes, CSEM and Non-CSEM. For the Non-CSEM class, we gathered $3,031,802$ unique paths for dumps of eight computer machines that host Non-CSEM files and $2,864,105$ for the CSEM class that was provided to us by the Spanish LEA. 
After pre-processing these paths, we ended with $2,065,590$ unique instances distributed as $924,445$ and $1,141,145$ for the CSEM and the Non-CSEM classes, respectively.

\section{Empirical Evaluation}
\label{sec: eval}

\subsection{Experimental Setting}
\label{sec: exp}
The experiments were carried out on a PC with an Intel(R) Core(TM) i7 processor with $32$ GB of RAM under Windows-10. We used Python3 with Scikit-Learn\footnote{\url{https://scikit-learn.org/stable/}} for implementing the classifiers.

Regarding the File Name Classifier's configuration, we used character n-grams, extracting patterns from two to five grams \cite{alnabki2020filenameclassification}. Also, we set thresholds for the minimum and the maximum gram proportion to $0.999$ and $0.0005$, respectively. For the LR classifier, we set the parameter $C$ to $100$, empirically, which refers to the inverse of regularization strength, and we activate the class weight parameter to consider the imbalance of the classes while training. The rest of the parameters were left to their default values, as the Scikit-Learn library set them. The File Path Classifier used the configurations as the File Names Classifier.

To estimate the models' performance, we report the performance of each classifier on a test set.
For the File Name Classifier, the dataset has $890,000$ samples before pre-processing and we split by $80/20$ for the training and the testing sets, respectively. Detailed description of the dataset size information is given in Table \ref {tab:dataset_fn}.

\begin{table}[htp]
\caption{Description of the used dataset to train the FNC.}
\centering
\label{tab:dataset_fn}
\begin{tabular}{lrrr}
\hline
 & \multicolumn{1}{l}{\textbf{CSEM}} & \multicolumn{1}{l}{\textbf{Non-CSEM}} & \multicolumn{1}{l}{\textbf{Total}} \\ \hline
Before pre-processing & 90,000 & 800,000 & 890,000 \\
Training set size & 72,000 & 640,000 & 712,000 \\
Testing set size & 18,000 & 160,000 & 178,000 \\ \hline
After pre-processing & 37,648 & 537,807 & 575,455 \\
Training set size & 30,161 & 421,175 & 451,336 \\
Testing set size & 7,487 & 116,632 & 124,119 \\ \hline
\end{tabular}
\end{table}

Unlike the file name samples, we could not split the training and the testing set on a fixed percentage, and this is because the samples of these sets must be non-intersected. Hence, the machines used to obtain the file paths' dump are distinct. Table \ref{tab:dataset_pn} gives detailed information about the dataset class size.

\begin{table}[htp]
\centering
\caption{Description of the used dataset to train the FPC}
\label{tab:dataset_pn}
\centering
\begin{tabular}{llll}
\hline
 & \textbf{CSEM} & \textbf{Non-CSEM} & \textbf{Total} \\ \hline
Before pre-processing & 2,864,105 & 3,031,802 & 5,895,907 \\
Training set size & 1,971,585 & 1,971,585 & 3,943,170 \\
Testing set size & 892,520 & 1,060,217 & 1,952,737 \\ \hline
After pre-processing & 924,445 & 1,141,145 & 2,065,590 \\
Training set size & 31,925 & 80,928 & 112,853 \\
Testing set size & 892,520 & 1,060,217 & 1,952,737 \\ \hline
\end{tabular}
\end{table}

Finally, to test the performance of both file name and file path models, we created a binary dataset of $50,000$ samples, equally distributed between the classes. We sampled $50,000$ file paths and another $50,000$ file names of the test sets of the path names and the file names sets randomly, respectively. Then, we created a balanced synthesized test set by fusing these two sets. A sample is considered CSEM if its name or path were sampled from a CSEM instance; otherwise, it is tagged as Non-CSEM. 

\subsection{Evaluation Metric}
\label{sec: metric}
The principal objective of this work is to assist LEAs in detecting CSEM through their file names, avoiding the exposure of an agent to CSEM. Therefore, it is desirable to have a low number of false negatives - a file named with CSEM content identified as a Non-CSEM - than a low number of false positives, i.e., Non-CSEM file name wrongly categorized as a CSEM. Hence, it is desirable to obtain a high recall of the CSEM class rather than the Non-CSEM class.

\textit{Recall} metric for a class is calculated as the total number of samples correctly classified for that class (the True Positives \textit{TP}), over the total number of samples of that class (the True Positives \textit{TP} and the False Negatives \textit{FN}). Equation (\ref{eq:recall_eq}) shows how Recall is estimated for a given class.
\begin{equation}
\label{eq:recall_eq}
Recall= \frac{TP}{TP + FN}.
\end{equation}

Nevertheless, the \textit{precision} of a classifier is also a crucial factor in measuring its performance, as it shows the proportion of correctly identified samples. Class precision is calculated as a ratio of correctly classified file names of that class (the True Positives \textit{TP}) to the total number of predicted positive samples of that class (the True Positives \textit{TP} and the False Positives \textit{FP}), and it is given in Equation (\ref{eq:precision_eq}).
\begin{equation}
\label{eq:precision_eq}
Precision= \frac{TP}{TP + FP}.
\end{equation}

Finally, the F1 score of a class summarizes the two before-mentioned metrics as it refers to the harmonic mean of the precision and recall and it is calculated following to Equation (\ref{eq:f1}).
\begin{equation}
\label{eq:f1}
F1 = \frac{ 2*(Recall * Precision)}{(Recall + Precision)}.
\end{equation}

Additionally, it has been proved that the accuracy metric is not reliable when the dataset is imbalanced \cite{chen2017harnessing}, as in our case, where the majority of the samples are Non-CSEM file names. An alternative metric is to use \textit{average class recall}, rather than using overall dataset level accuracy.

\subsection{Empirical Results}
In this section, we evaluate both classifiers and the proposed fusion methods, as described in Section \ref{sec: methodology}.

Table \ref{table:results_fnc} analyzes the impact of these features used to boot the file name representation. Our results show that when all the representation techniques are joined, we could obtain the best classification performance for the FNC with an average class recall of $0.98$ and an F1 score of $0.96$.

\begin{table}[htp]
\centering
\caption{The impact of file name representation on the performance of the FNC.}
\label{table:results_fnc}
\resizebox{\linewidth}{!}{
\begin{tabular}{clrrrr}
\hline
\textbf{\begin{tabular}[c]{@{}c@{}}Pre-processing \\ stages\end{tabular}} & \multicolumn{1}{c}{\textbf{Category}} & \multicolumn{1}{c}{\textbf{Precision}} & \multicolumn{1}{c}{\textbf{Recall}} & \multicolumn{1}{c}{\textbf{F1}} & \multicolumn{1}{c}{\textbf{Support}} \\ \hline
\multirow{3}{*}{\textbf{\begin{tabular}[c]{@{}c@{}}Pure file \\ name\end{tabular}}} & CSEM & 0.99 & 0.99 & 0.99 & 150,448 \\
 & Non-CSEM & 0.84 & 0.95 & 0.89 & 7,510 \\
 & Avg & 0.92 & 0.97 & 0.94 & 157,958 \\ \hline
\multirow{3}{*}{\textbf{\begin{tabular}[c]{@{}c@{}}Digit \& \\ special char.\end{tabular}}} & CSEM & 0.99 & 0.99 & 0.99 & 113,855 \\
 & Non-CSEM & 0.86 & 0.95 & 0.90 & 7,484 \\
 & Avg & 0.93 & 0.97 & 0.94 & 121,339 \\ \hline
\multirow{3}{*}{\textbf{Binary representation}} & CSEM & 0.99 & 0.99 & 0.99 & 114,162 \\
 & Non-CSEM & 0.89 & 0.95 & 0.92 & 7,484 \\
 & Avg & 0.94 & 0.97 & 0.95 & 121,646 \\ \hline
\multirow{3}{*}{\textbf{\begin{tabular}[c]{@{}c@{}}Orthographic \\ representation\end{tabular}}} & CSEM & 0.99 & 0.99 & 0.99 & 116,632 \\
 & Non-CSEM & 0.90 & 0.97 & 0.93 & 7,487 \\
 & Avg & 0.95 & 0.98 & 0.96 & 124,119 \\ \hline
\end{tabular}
}
\end{table}

Afterward, we evaluated the FPC on its test set, as shown in Table \ref{table:results_FPC}. The FPC obtained $0.97$ for both of the average class recall and the F1 score, which is slightly higher than the FNC, which scored $0.98$ and $0.96$, respectively.

\begin{table}[htp]
\centering
\caption{The performance of the FNC and the FPC classifiers on their test sets.}
\label{table:results_FPC}
\resizebox{\linewidth}{!}{
\begin{tabular}{ccrrrr}
\hline
\textbf{Model Name} & \textbf{Category} & \multicolumn{1}{c}{\textbf{Precision}} & \multicolumn{1}{c}{\textbf{Recall}} & \multicolumn{1}{c}{\textbf{F1}} & \multicolumn{1}{l}{\textbf{Support}} \\ \hline
\multicolumn{1}{c|}{\multirow{3}{*}{\textbf{FNC}}} & Non-CSEM & 0.99 & 0.99 & 0.99 & 116,632 \\
\multicolumn{1}{c|}{} & CSEM & 0.90 & 0.97 & 0.93 & 7,487 \\
\multicolumn{1}{c|}{} & Average & 0.95 & 0.98 & 0.96 & 124,119 \\ \hline
\multicolumn{1}{c|}{\multirow{3}{*}{\textbf{FPC}}} & Non-CSEM & 0.96 & 1.00 & 0.98 & 1,060,217 \\
\multicolumn{1}{c|}{} & CSEM & 1.00 & 0.95 & 0.97 & 892,520 \\
\multicolumn{1}{c|}{} & Average & 0.97 & 0.97 & 0.97 & 1,952,737 \\ \hline
\end{tabular}
}
\end{table}

In addition to reporting the performance of each classifier individually, we analyze two techniques of fusing them, as described earlier. Table \ref{table:results_fusing} shows that using two standalone classifiers, one for the file path and one for the file name, surpasses the single iterative classifier approach. The two classifiers architecture could achieve an average class recall of $0.98$, which is higher than the other approach that iteratively uses the FNC and scores $0.74$.

\begin{table}[htp]
\centering
\caption{A comparison between two techniques of fusing the FPC and FNC models. The values in bold refer to the best prediction F1 score.}
\centering
\label{table:results_fusing}
\resizebox{\linewidth}{!}{
\begin{tabular}{ccrrrr}
\hline
\textbf{\begin{tabular}[c]{@{}c@{}}Classifiers \\ Fusion method\end{tabular}} & \textbf{Category} & \multicolumn{1}{c}{\textbf{Precision}} & \multicolumn{1}{c}{\textbf{Recall}} & \multicolumn{1}{c}{\textbf{F1}} & \multicolumn{1}{l}{\textbf{Support}} \\ \hline
\multicolumn{1}{c|}{\multirow{3}{*}{\textbf{\begin{tabular}[c]{@{}c@{}}Two Standalone \\ Classifiers\end{tabular}}}} & Non-CSEM & 0.99 & 0.98 & 0.98 & 25000 \\
\multicolumn{1}{c|}{} & CSEM & 0.98 & 0.99 & 0.98 & 25000 \\
\multicolumn{1}{c|}{} & Average & \textbf{0.98} & \textbf{0.98} & \textbf{0.98} & 50,000 \\ \hline
\multicolumn{1}{c|}{\multirow{3}{*}{\textbf{\begin{tabular}[c]{@{}c@{}}Single Iterative \\ Classifier\\ Approach\end{tabular}}}} & Non-CSEM & 0.68 & 0.94 & 0.79 & 25000 \\
\multicolumn{1}{c|}{} & CSEM & 0.90 & 0.55 & 0.68 & 25000 \\
\multicolumn{1}{c|}{} & Average & 0.73 & 0.74 & 0.78 & 50,000 \\ \hline
\end{tabular}
}
\end{table}

\section{Conclusions and Future Work}
\label{sec: conclusion}
In this paper, we presented a supervised machine learning approach to identify files that may contain Child Sexual Abuse Material (CSEM) from regular files (Non-CSEM). Given that this work aims to build a fast CSEM prediction, only file names and paths are used. We proposed two solutions: 1) building two standalone classifiers: a File Name Classifier (FNC) and File Path Classifier (FPC), and then fusing their outputs into a single decision, and 2) dividing the file path into a list of folder names and using the FNC to classify each name in the path. Our results strengthen the superiority of the former approach as it obtained an average class recall of $0.98$, while the latter scored an average class recall of $0.74$.
 
For the FNC, we pre-processed the text and boosted it with two features: binary and orthography, which increased the recall rate of the CSEM class from $0.89$ to $0.93$ and scored an average class recall of $0.98$. 
Regarding the FPC, it used similar architecture to the FNC, but it differs in the pre-processing stage, and it achieved an average class recall of $0.97$.

The empirical evaluation was conducted on a dataset extracted from the file names and file paths. As future work, we are looking forward to enlarging the dataset by obtaining samples from various seized computers, allowing the model to be exposed to wider CSEM file names patterns. Furthermore, once the dataset is extended, we aim to build a character-based language model \cite{conneau2019cross} for CSEM files. The assessment of transformer-based models, such as BERT \cite{luo2018active}, and XLNet \cite{Zhilin2019XLNet} for text classification is part of our immediate future research, as they have shown promising results on various NLP tasks.

\section*{ACKNOWLEDGEMENTS}
This work was supported by the framework agreement between the University of Le\'on and INCIBE (Spanish National Cybersecurity Institute) under Addendum 01. This research has been funded with support from the European Commission under the 4NSEEK project with Grant Agreement 821966. This publication reflects the views only of the author, and the European Commission cannot be held responsible for any use which may be made of the information contained therein.


\bibliography{mybibfile}

\end{document}